\documentclass[12pt]{article}
\usepackage{amsmath,amsfonts}
\advance \textheight by \topskip
\makeatletter
\@addtoreset{equation}{section}
 \def\theequation{\thesection.\arabic{equation}}
\makeatother

\newcommand{\beq}{\begin{equation}}
\newcommand{\bee}{\end{equation}}
\newcommand{\beqa}{\begin{eqnarray}}
\newcommand{\eeqa}{\end{eqnarray}}

\def\>{\rangle}
\def\<{\langle}

\begin{document}

\title{
\begin{flushright}
{\small PM/03-10}\\[1cm]
\end{flushright}
{\bf Field Theoretic Realizations for Cubic Supersymmetry}}


\author{
{\sf   N. Mohammedi} \thanks{e-mail:
nouri@celfi.phys.univ-tours.fr}$\,\,$${}^{a},$
{\sf  G. Moultaka }\thanks{e-mail:
moultaka@lpm.univ-montp2.fr}$\,\,$${}^{b}$
 and
{\sf M.~Rausch de Traubenberg}\thanks{e-mail:
rausch@lpt1.u-strasbg.fr}$\,\,$$^{c}$\\
\\
{\small ${}^{a}${\it Laboratoire de Math\'ematiques et Physique Th\'eorique,
Universit\'e Fran\c{c}ois Rabelais,}} \\
{\small {\it Facult\'e des Sciences et Techniques,
Parc de Grandmont, F-37200 Tours, France.}}  \\ 
{\small ${}^{b}${\it Laboratoire de Physique 
Math\'ematique et Th\'eorique, CNRS UMR 5825, 
Universit\'e Montpellier II,}}\\
{\small {\it Place E. Bataillon, 34095 Montpellier,
France}}\\
{\small ${}^{c}${\it
Laboratoire de Physique Th\'eorique, CNRS UMR  7085,
Universit\'e Louis Pasteur}}\\
{\small {\it  3 rue de
l'Universit\'e, 67084 Strasbourg, France}}}
\date{}
\maketitle
\vskip-1.5cm

\vspace{2truecm}

\begin{abstract}

\noindent
We consider a four dimensional space-time symmetry which is a non trivial 
extension of the Poincar\'e algebra, different from supersymmetry and not 
contradicting  {\sl a priori} the well-known no-go theorems.
We investigate some field theoretical aspects of this new symmetry and
construct invariant actions for non-interacting fermion and 
non-interacting boson multiplets. 
In the case of the bosonic multiplet, where two-form fields
appear naturally, we find that this symmetry is compatible with a  local 
$U(1)$ gauge symmetry,  
only when the latter is gauge fixed by a `t Hooft-Feynman term.   

\end{abstract}

\vspace{2cm}
PACS numbers: 03.50.Kk,   03.65.Fd, 11.30.Ly 

keywords: generalization of supersymmetry, Field Theory, 
algebraic methods, Lie (super)algebras.

\newpage
\section{Introduction}
\renewcommand{\theequation}{1.\arabic{equation}}   
\setcounter{equation}{0}

Over several decades, supersymmetry (SUSY) \cite{susy1,susy2, susy3} 
has gradually gained the status of a 
cornerstone in the search for a unified description of elementary particle 
physics, and more generally of the four fundamental interactions.
Despite the present absence of any direct SUSY signatures,
some indirect experimental evidence together with a wealth of theoretically 
appealing features can be embedded in the so-called 
minimal supersymmetric standard model (MSSM) and its constrained as well
as extended versions (see {\it e.g.} \cite{nilles}), thus sharing the 
tremendous experimental success of the Standard Model of particle physics and 
predicting new physics  at the $O(1)TeV$ energy scale.  
Moreover, from a purely algebraic point of view, the consideration of
supersymmetric theories found its mathematical insight in some
extensions of Lie algebras called Lie superalgebras, in particular,
evading the conditions of validity of the Coleman-Mandula theorem \cite{cm}
and leading to a non-trivial extension of the Poincar\'e algebra \cite{hls}.
In addition, the introduction
of Lie superalgebras has lead to new powerful mathematical tools.

In supersymmetric theories, the extensions of the Poincar\'e
algebra are obtained from a ``square root'' of the translations,
``$QQ \sim P$''.
It is tempting to consider other alternatives where the new algebra
is obtained from yet higher order roots.
The simplest alternative which we will consider in this paper is 
``$QQQ \sim P$''. It is important to stress that such structures
are not Lie (super)algebras  (even though they  contain a Lie sub-algebra),
and as such escape {\it a priori} the Coleman-Mandula \cite{cm}
as well as  the Haag-Lopuszanski-Sohnius no-go theorems \cite{hls}.
Furthermore, as far as we know, no no-go theorem associated with such types
of extensions has been considered in the literature. 
This can open interesting  possibilities to search for a field theoretic
realization of a non trivial extension of the Poincar\'e algebra which 
{\it is not the supersymmetric one}. If successful, this might throw
a new light on how to construct physical models.\\ 
 

Regarding the algebra {\it per se}, several possibilities have already been
considered in the literature. Here we are  focusing on one
of the possible extensions called fractional supersymmetry (FSUSY),
\cite{fsusy} -- \cite{flie2}.  
Basically, in such extensions,
 the generators of the Poincar\'e algebra are obtained as   $F-$fold 
symmetric products of more fundamental generators, leading to the
``$F^{\mathrm{th}}-$root'' of translation: ``$Q^F \sim P$'' with $F$
a positive integer. The $F-$Lie algebras, the structures which underlie FSUSY,
 are defined in \cite{flie1, flie2}  
 in full analogy with SUSY and its underlying  Lie superalgebra structure.\\

The aim of this paper is to provide the first field theoretic construction in 
$(1+3)$ dimensions of an FSUSY with $F=3$, which we will 
refer to as {\it cubic supersymmetry} 3SUSY. For this purpose we will first 
determine finite dimensional fermion and boson multiplet representations of 
the 3SUSY.
Then, we will seek  explicit 3SUSY invariant actions and discuss their
specific features. We find that the fermion multiplets are made of three
definite chirality fermions which are degenerate in mass, while the boson 
multiplets contain Lorentz scalars, vectors and two-forms. 
It turns out that cubic SUSY forbids canonical 
kinetic terms for fermion fields and leads either to higher order derivative 
ones or requires interaction with some constant background fields. A striking
feature for the boson multiplets is the compatibility of 3SUSY with gauge
symmetry only when the latter is gauge fixed in the usual way. We provide also
a short discussion of the Noether currents. The rest of the paper is organized
as follows: in section 2 we give the cubic supersymmetry algebra and 
its finite dimensional representations. In section 3
we construct the fermion multiplets and discuss the corresponding allowed
actions. Section 4 is devoted to the boson multiplets and to the 
construction of gauge actions compatible with 3SUSY. Noether currents
are briefly discussed in section 5 where some extra comments are made.
The conclusions and outlook are summarized in section 6.
More extended material
for the algebraic construction of 3SUSY can be found in appendix A.
Appendix B contains some notations and conventions.        


\section{Non trivial extensions of the Poincar\'e algebra}
\renewcommand{\theequation}{2.\arabic{equation}}   
\setcounter{equation}{0}

The FSUSY algebra is generated by the usual  generators of the Poincar\'e
algebra together with additional ``supercharges''.
These new
generators have to be in some representation of the Lorentz algebra. 
Historically, it was firstly believed that FSUSY could only apply in low
dimensional systems ($D \le 1+2$),
where representations which are neither bosonic nor
fermionic exist \cite{fr, prs, fsusy2d, fsusy3d}. 
A next step in the understanding of FSUSY was achieved by the discovery
that FSUSY could be extended to any space time-dimension \cite{flie1},
considering infinite dimensional representations of the Lorentz algebra
(fractional spin).
However, these representations being not exponentiable
(they are representations of the Lie algebra but 
not representations of the Lie group), all the results
are valid at the level of the Lie algebra, and consequently the
principle of equivalence of special relativity is lost. 
These results were looking like a breakthrough for the construction
of FSUSY in dimension higher than three.
In the meantime,  it was understood that finite dimensional
$F-$Lie algebras ({\it i.e.} involving representations which are not
infinite dimensional and which are exponentiable) 
could be obtained  by an 
inductive process
starting from any simple Lie (super)algebra \cite{flie2}. 
Among these families of
examples, it was observed that, under an In\"on\"u-Wigner contraction
of some of the $F-$Lie algebras,
non-trivial extensions of the Poincar\'e algebra (which are 
not the usual supersymmetric ones) can be obtained.

In this paper, we will be interested in one of these extensions, the 3SUSY. 
It is constructed from the Poincar\'e generators
(in any space-time dimension) $L_{mn}$
and $P_m$ together with some  additional supercharges $Q_m$ in the vector representation,
 satisfying the trilinear relations

\beqa
\label{ext-susy}
Q_m Q_n Q_r + Q_m Q_r Q_n& +& Q_n Q_m Q_r + Q_n Q_r Q_m + Q_r Q_m Q_n +
Q_r Q_n Q_m   \nonumber \\
&=&\eta_{mn} P_r + \eta_{nr} P_m +\eta_{mr} P_n, 
\eeqa

\noindent
with $\eta_{mn}$ the Minkowski metric.
This algebra can be compared in some sense with the algebraic extension
studied in  \cite{ker, wt}.

The algebraic  features of this new structure is summarized
in  appendix A and leads to the following algebraic extension
of the Poincar\'e algebra

\beqa
\label{F-poin}
&&\left[L_{mn}, L_{pq}\right]=
\eta_{nq} L_{pm}-\eta_{mq} L_{pn} + \eta_{np}L_{mq}-\eta_{mp} L_{nq},
\ \left[L_{mn}, P_p \right]= \eta_{np} P_m -\eta_{mp} P_n, \nonumber \\
&&\left[L_{mn}, Q_p \right]= \eta_{np} Q_m -\eta_{mp} Q_n, \ \
\left[P_{m}, Q_n \right]= 0, \\
&&\left\{Q_m, Q_n, Q_r \right \}=
\eta_{m n} P_r +  \eta_{m r} P_n + \eta_{r n} P_m, \nonumber
\eeqa

\noindent
where $\{Q,Q,Q \}$  stands for the symmetric product of order $3$
(as defined in (\ref{ext-susy})). 
It has to be  emphasized that such an extension is not  the 
parasupersymmetric one considered in \cite{ni}, even if it involves
also a trilinear bracket.\\ 

The aim of the present paper is to try to
implement this new structure in a field theoretical setting.
As noted in appendix A, we have found a twelve-dimensional matrix 
representation of the algebra (\ref{F-poin})

\beqa
\label{Mat12}
Q_m = \begin{pmatrix}0&\Lambda^{1/3}\gamma_m&0 \cr
                 0&0& \Lambda^{1/3} \gamma_m \cr
                 \Lambda^{-2/3} P_m&0&0 \end{pmatrix} , & 
\eeqa 

\noindent
with $\Lambda$ a parameter with mass dimension and $\gamma_m$ the
$4D$ Dirac matrices.
The $Q$'s being in the vector representation of the Lorentz algebra,
we also have (see appendix A)

\beqa
\label{Lorentz}
J_{mn} = \frac{1}{4}\left(\gamma_m \gamma_n - \gamma_n \gamma_m\right)+
i(x_m P_n -x_n P_m),\ \
[J_{m n}  ,Q_r  ]  = \eta_{n r} Q_n - \eta_{m r} Q_m,
\eeqa

\noindent
with $P_m= -i \partial_m$.
However, this  representation is reducible,
leading to the two inequivalent $6-$dimensional representations:

\beqa
\label{Mat6}
Q_m = \begin{pmatrix}0&\Lambda^{1/3}\sigma_m&0 \cr
                 0&0& \Lambda^{1/3} \bar \sigma_m \cr
                 \Lambda^{-2/3} P_m&0&0 \end{pmatrix}, 
\eeqa 

\noindent
and

\beqa
\label{Mat6cc}
Q_m = \begin{pmatrix}0&\Lambda^{1/3}\bar \sigma_m&0 \cr
                 0&0& \Lambda^{1/3}  \sigma_m \cr
                 \Lambda^{-2/3} P_m&0&0 \end{pmatrix}.
\eeqa

\noindent
See  appendix B for the conventions for the $\sigma-$matrices
 and some useful relations.
These two representations are $CPT$ conjugate
of each other.   
As already mentioned,  the two representations (\ref{Mat6})
and (\ref{Mat6cc}) are certainly not the only possible  ones. 
It might as well be
that other matrix representations could
also lead  to some  interesting physical results.

We would like to end up this section by some comparison with
the ordinary supersymmetric extension of the Poincar\'e algebra.
Recall that if one tries to construct representations of the SUSY
algebra by
considering Clifford algebra of polynomial
(with $8$ variables), one  ends up with $16 \times 16$ matrices. Introducing
the Grassmann algebra ({\it i.e.} the 
$\theta_\alpha, \bar \theta_{\dot \alpha}$ and their derivative 
$\partial_{\theta_\alpha},  \bar \partial_{\bar \theta_{\dot \alpha}}$) 
in matrix representation, these
matrices reduce then to the
supercharges in the superspace language. Moreover, the number of variables
being twice as  many as the number of independent variables, leads  automatically to
reducible representations in the superspace approach. 
However, within this approach,  the matrix representation can be forgotten
and only the algebra between the $\theta$'s and the $\partial_\theta$ is
useful.
It is indeed also possible, for FSUSY, to introduce 
some additional variable, 
$\eta^m =\begin{pmatrix} 0&\sigma^m&0 \cr 0&0&\bar \sigma^m\cr 0&0&0  
\end{pmatrix}$ and $\partial_{\eta_m} = \begin{pmatrix} 
0&0&0\cr \bar \sigma_m \cr 0&\sigma_m& 0\end{pmatrix}$ such that 
we have $Q_m = \partial_{\eta^m} + f_{mnp}{}^q \left(\eta^n \eta^p + 
\eta^p \eta^n\right) P_q$   with  $f_{mnr}{}^{ s}= \eta_{m n} \delta_r{}^{ s} +  \eta_{m r} \delta_n{}^{ s} + 
\eta_{r n} \delta_m{}^{ s}$.
 However, we were not able within these variables to introduce some
adapted version for a superspace. We will then continue with the matrices
(\ref{Mat6}) and (\ref{Mat6cc}).

\section{Fermion multiplets}
\renewcommand{\theequation}{3.\arabic{equation}}   
\setcounter{equation}{0} 

In this section we construct an invariant action under the 3SUSY algebra
(\ref{F-poin}) and the representations (\ref{Mat6}) and (\ref{Mat6cc}).
As usual, the content of the representation is not only specified by
the form of the matrix representation, but also by the behavior of the
vacuum under Lorentz transformations. If we denote $\Omega$ the vacuum,
which is in some specified representation of the Lorentz algebra,
with $\Sigma_{mn}$ the corresponding Lorentz generators, then 
$J_{mn}$ given in (\ref{Lorentz}) is  replaced by
$J_{mn} + \Sigma_{mn}$. 
In the  case,  where $\Omega$ is a Lorentz scalar,
one sees that the  multiplet of 
representation (\ref{Mat6}) contains two left-handed  and one
 right-handed fermions, while the multiplet of the representation 
(\ref{Mat6cc}) contains one left-handed  and two right-handed fermions.
These two multiplets are $CPT$ conjugate. 
At first sight, it might  seem surprising that a multiplet contains fields
of the same statistics (here only fermions). Indeed, this comes from the
fact that we are considering a supercharge $Q$ in the vector representation,
in order to extract the ``cubic root of the translation''. In supersymmetric
theories the square root is extracted using spinors, and consequently
representations of SUSY contain both fermions and bosons. 
If we would expect something similar in 3SUSY, the $Q$ have to be in
the spin $1/3$ representation of the Lorentz algebra (see \cite{flie1}).
But this representation is firstly infinite dimensional, and secondly
cannot be exponentiated
(see {\it e.g.} \cite{kpr}). Therefore, it does not define a representation of the Lie
group  $\overline{SO(1,3)}$ 
(in $(1+2)$ dimensions, however, a similar extension is
possible \cite{fsusy3d} and   such  representations describe relativistic
anyons \cite{anyon}.)

Consider the multiplet associated with the matrices (\ref{Mat6}). If we denote
${\mathbf \Psi}= \begin{pmatrix} \psi_{1 \alpha} \cr \bar \psi_2^{\dot \alpha}
\cr  \psi_{3 \alpha} \end{pmatrix}$, then under a 3SUSY 
transformation we have  $\delta_\varepsilon {\mathbf \Psi}
=\varepsilon^m Q_m {\mathbf \Psi}$ and 
we obtain (see appendix B for spinor notations) 

\beqa
\label{transfo1}
\delta_\varepsilon \psi_{1 }& =& \varepsilon^n 
\Lambda^{1/3}\sigma_{n }
\bar \psi_2 \nonumber \\
\delta_\varepsilon \bar \psi_{2}& =& \varepsilon^n \Lambda^{1/3}
\bar \sigma_{n}
\psi_{3 } \\
\delta_\varepsilon \psi_{3}& =& \varepsilon^n 
\Lambda^{-2/3}P_n \psi_{1} \nonumber
\eeqa

\noindent
Using $\Big(\sigma_m{}_{\alpha \dot \beta}\Big)^\star=  
\sigma_m{}_{\dot \beta \alpha}$ and the relations in  appendix B, 
we find the following transformations for  the
conjugate fields

\beqa
\label{transfo1cc}
\delta_\varepsilon \bar \psi_{1}& =& 
\varepsilon^n \Lambda^{1/3}\bar \sigma_{n}
\psi_{2} \nonumber \\
\delta_\varepsilon  \psi_{2}& =& \varepsilon^n \Lambda^{1/3}
\sigma_{n}
\bar \psi_{3} \\
\delta_\varepsilon \bar \psi_{3}& =& 
\varepsilon^n \Lambda^{-2/3}P_n \bar \psi_{1} \nonumber
\eeqa

\noindent
with $\varepsilon^n$, the parameter of the  3SUSY transformation, 
a purely imaginary number with mass dimension $-1/3$.
At this point it can be observed that if one considers $4D$ Majorana spinors
$\psi_i=\begin{pmatrix} \psi_{i \alpha} \cr \bar \psi_i{}^{\dot \alpha}
\end{pmatrix}$,
instead of $2D$ Weyl spinors,
 the matrices (\ref{Mat12}) ensure that
$\delta_\varepsilon \psi_i$ are also Majorana
(as can be explicitly seen in the appendix B). This means that an invariant
3SUSY action can also be constructed with $4D$ Majorana spinors.
  In the sequel we will consider both representations.

Finally, notice  that one can associate a grade (or degree) to each of the
fermions in the following manner: 
$\psi_1$ is of grade
$-1$, $\psi_2$ of grade $0$ and $\psi_3$ of grade $1$.
Then $Q$ turns out to be of grade $-1$ 
and the transformations properties (\ref{transfo1}) and (\ref{transfo1cc})
are compatible with this grading: $\psi_1  \buildrel{\hbox{Q}} \over
\longrightarrow
\psi_3 \buildrel{\hbox{Q}} \over \longrightarrow
\psi_2 \buildrel{\hbox{Q}} \over \longrightarrow \psi_1
$. Moreover, if one wants that both sides of (\ref{transfo1}) and 
(\ref{transfo1cc}) have the same grade, then 
one has to assign a grade $-1$ to the parameter 
$\varepsilon$.

We will try  now to construct 3SUSY invariant Lagrangians up to a surface
term. That is, we will seek a Lagrangian ${\cal L}$ such that 

$$\delta_\varepsilon {\cal L} = \varepsilon_m \partial^m \left(\cdots
\right).$$

\noindent
Since, under 3SUSY    only $\psi_3$ transforms as a total
derivative ((\ref{transfo1}),  (\ref{transfo1cc})), and given the 
grading structure,
the only possible candidates bilinear in the  fermion fields involve couplings of
either $\psi_1$ to $\psi_3$ or  $\psi_2$ to itself.
However, this is not sufficient.
For instance, one can easily show that the simplest kinetic  term

\beqa
\label{L1}
{\cal L}=
\left(\psi_1 i \sigma^m \partial_m \bar \psi_3  +
\psi_3 i \sigma^m \partial_m \bar \psi_1\right)
 +\left(\bar \psi_1  i\bar  \sigma^m \partial_m \psi_3 +
 \bar \psi_3  i \bar \sigma^m \partial_m \psi_1\right) +
 \psi_2  i \sigma^m \partial_m \bar \psi_2 +
\bar \psi_2  i\bar \sigma^m \partial_m \psi_2
 \nonumber \\
\eeqa

\noindent
does not transform as a surface term. The reason for this  being that 
the Pauli matrices do not commute. This means that unconventional
kinetic terms have to be considered. However, as we will see, 
conventional mass terms are still allowed. 
More generally, let us consider
the following Lagrangian (in the $4D$ formalism for notational
convenience),

\beqa
{\cal L} = \bar \psi_1 {\cal O} \psi_3 + \bar \psi_3 {\cal O} \psi_1
 + \bar \psi_2 {\cal O} \psi_2
\eeqa

\noindent
where ${\cal O}$ is a $4 \times 4$ hermitian matrix operator
and $\psi_i$ are three Majorana spinors transforming as in (B.15). 
The variation of this Lagrangian under 3SUSY gives

\beqa
\delta_\varepsilon {\cal L} &=& \varepsilon^n 
\{ -\bar \psi_2  \gamma_n {\cal O}
\psi_3 + \bar \psi_1  {\cal O} P_n \psi_1 \} \nonumber \\
 &+ &\varepsilon^n \{ P_n \bar \psi_1  {\cal O} \psi_1 + \bar \psi_3  {\cal O}
\gamma_n \psi_2 \}  \\
 &+ &\varepsilon^n \{ - \bar \psi_3  \gamma_n {\cal O} \psi_2 +
\bar \psi_2  {\cal O} \gamma_n \psi_3 \} \nonumber 
\eeqa

\noindent
For $\delta_\varepsilon {\cal L}$ to be a surface term, the operator ${\cal O}$
should fulfill

\beqa
\label{conditions}
[{\cal O}, \partial_n ] = [{\cal O}, \gamma_n ] = 0 
\eeqa
  
\noindent
The most general form compatible with (\ref{conditions}) is found to be

\beqa
\label{nouri}
{\cal O} = \begin{pmatrix}  [m + c^n \partial_n] \times I & 0 \cr
                     0 & [m + c^n \partial_n] \times I  \end{pmatrix}
\eeqa

\noindent
where $I$ is the two by two identity matrix, $m$ is a (constant) mass and
$c^m$ a Lorentz vector  operator commuting with $ \partial_n$. 

If the $\psi$'s are Majorana spinor fields then,
$c_m$ should contain a derivative 
 ($c^m = \frac{\partial^m}{\Lambda}
f(\frac{\Box}{\Lambda^2}) $) otherwise the derivative part of 
${\cal L}$ corresponding to $\psi_1, \psi_2$ reduces to a surface term. 
The simplest 3SUSY  Lagrangian reads (when $f$ is the identity 
function) 

\beqa
{\cal L}_1 = \bar \psi_1  \frac{\Box}{\Lambda} \psi_3 +  
            \bar \psi_3  \frac{\Box}{\Lambda} \psi_1 +
           \bar \psi_2  \frac{\Box}{\Lambda} \psi_2 +
            m ( \bar \psi_1  \psi_3 +   \bar \psi_3  \psi_1 
            +  \bar \psi_2  \psi_2 )
\eeqa

\noindent
Note that for the kinetic term we used the natural mass scale appearing
in the representation of the 
algebra (\ref{Mat12}) while $m$ is an additional mass parameter.
Some remarks are in order here. It is straightforward to redefine
the fields $ \psi_1, \psi_3$ in terms of positive squared mass ($m \Lambda$)
 eigenstates. However, the classical equation of motion from ${\cal L}_1$
leads only to a Klein-Gordon type equation, thus determining the mass 
eigenvalue but not the spin content (the equation corresponding to the
Pauli-Lubanski Casimir is missing). This means that the non-interacting 
$\psi$ fields behave like ghosts (anti-commuting spin zero fields). 
The spinorial character could then be restored by the inclusion of interaction 
terms with additional fields. 



%
%


\bigskip
There is another possibility to construct invariant actions involving
fermions. This is achieved by introducing an extra fermionic  multiplet 

\beqa
{\bf \Lambda}^n = {\bf \Psi} \otimes \Omega^n = 
\begin{pmatrix} \lambda_1^n \cr
                \lambda_2^n \cr
                \lambda_3^n  \end{pmatrix} 
\eeqa

\noindent
where ${\bf \Psi}$ is a triplet of Majorana fermions transforming under
3SUSY as in (\ref{transfo1}, \ref{transfo1cc}) and $\Omega^n$ (the vacuum)
is a Lorentz vector and a 3SUSY singlet. Consequently, 
${\bf \Lambda}^n$ describes a triplet of Rarita-Schwinger fields transforming
under 3SUSY as spin $1/2$ fields (\ref{transfo1}, \ref{transfo1cc}), 
with an extra vector index attached ({\it e.g.} 
$\delta_\varepsilon \lambda_1^n = \varepsilon^m \gamma_m \lambda_2^n $). 
It is then straightforward to show that the Lagrangian

\beqa
{\cal L}_3 &=& \bar \psi_1  \partial_n \lambda_3^n + \bar \psi_3  \partial_n 
\lambda_1^n + \bar \psi_2  \partial_n \lambda_2^n 
   +  \partial_n  \bar \lambda_3^n  \psi_1 +   
         \partial_n  \bar \lambda_1^n  \psi_3 + 
         \partial_n  \bar \lambda_2^n  \psi_2 \nonumber \\
   &+&  m (\bar \psi_1  \psi_3 + \bar \psi_3  \psi_1 
            + \bar \psi_2  \psi_2) + 
        M (\bar \lambda_1^n  {\lambda_3}_n +  \bar \lambda_3^n  {\lambda_1}_n 
            + \bar \lambda_2^n  {\lambda_2}_n )  
\eeqa
\noindent
transforms as a surface term. Again, 
as in the previous Lagrangian, the kinetic part of the Rarita-Schwinger
is non-conventional. Here, Rarita-Schwinger just means we are considering
a spin $3/2$ representation of the Lorentz group.
Let us now project out
 the spin-half content of  the $\lambda^n_k$'s which will be the dynamical 
degrees of freedom.
We choose the following form for reasons which will be justified
{\sl a posteriori}.

\beqa
\label{generic}
\lambda_k^n = a_0^n \chi_0^{(k)} + a_1^n \chi_1^{(k)} + a_2^n \chi_2^{(k)} +
 i \gamma^n \chi_4^{(k)}
\eeqa

\noindent
where $a_0^n, a_1^n, a_2^n$ are real constant vectors which we assume, without
loss of generality, to be orthogonal to each other
and not light-like, and the $\chi^{(k)}$'s  
Majorana fields, so that the $\lambda_k^n$'s are real. 
Note first that if we keep only the term $i \gamma^n \chi_{4}^{(k)}$ in 
(\ref{generic}), then the Rarita-Schwinger field would be constrained by 
$\lambda^n = \frac14 \gamma^n \gamma_q \lambda^q$.
One can easily show that the latter constraint is not preserved by
3SUSY. It is thus important to keep enough terms in (\ref{generic}) so that
$\lambda^n_k$ are not constrained {\it a priori}. One way of doing this is   
to choose a sufficient number of non-zero vector (and/or two-form)
\footnote{This means that terms like $b^{mn} i \gamma_m \chi^{(k)},
b_n \gamma^{m n} \chi^{(k)}$, with $\chi$ a Majorana spinors
{\it etc.} could also be introduced.} components
so that there is a one-to-one correspondence between the $\chi$'s and
the $\lambda$'s.
%
In eq.(\ref{generic}) for each $k$, the four Majorana $\chi$'s have as many 
degrees of freedom as $\lambda^n$, that is  $16$ real components. To get a 
one-to-one correspondence, it is then sufficient to require that 
$a_0^n, a_1^n$ and $a_2^n$ be acolinear (we actually choose them orthogonal).
Then, the system expressing the components of $\lambda$ in terms of those of 
the four $\chi$'s is of rank $16$. In this specific case, 
using $\gamma^n= \frac{a_0^n}{a_0.a_0} {\not a_0} +
\frac{a_1^n}{a_1.a_1} {\not a_1} +\frac{a_2^n}{a_2.a_2} {\not a_2} +
\frac{a_2^n}{a_3.a_3} \not{a}_3$, with $a_3$ a  fourth vector orthogonal
to $a_0, a_1$ and $a_2$ and $\not a = a^n \gamma_n$
we get

\beqa
\label{chi}
\chi_{l}{}^{ (k)}&=& \frac{1}{a_l.a_l} 
\left(a_l.\lambda_k - i {\not a_l} \chi_4{}^{(k)}\right),
 \ \ 
l=0,1,2 \nonumber \\
\chi_4{}^{(k)} &=& -i \frac{\not{a_3}}{a_3.a_3} a_3. \lambda_k
\eeqa
This ansatz allows to determine unambiguously the 3SUSY transformation laws of 
the $\chi$'s and thus to obtain a 3SUSY invariant action involving only spin 
one-half fermions, with canonical kinetic terms and additional couplings
to background constant fields. Furthermore, one has to identify the  various 
physical fields as well as possible auxiliary ones in ${\cal L}_3$.
In general, the induced 3SUSY transformations for the $\chi_l$'s will depend
on the constant vectors $a_0^n, a_1^n, a_2^n$, thus making a physical 
interpretation rather difficult.

It is not difficult to obtain the 3SUSY transformation corresponding
to (\ref{chi})

\beqa
\label{rs}
\begin{array}{ll}
\left.
\begin{array}{ll}
k=1,2 &\left\{
\begin{array}{l}
\delta_\varepsilon \chi_l^{(k)}  =
\Lambda^{1/3}\left(
{\not \varepsilon} \chi_l^{(k+1)} + 
\left(2 i \frac{\varepsilon.a_l}{a_l.a_l} 
-2i \frac{\varepsilon.a_3}{a_3.a_3} \frac{\not a_l\not a_3}{a_l.a_l}\right)
\chi_4^{(k+1)}
 \right)  \cr
\delta_\varepsilon \chi_4^{(k)}  = \Lambda^{1/3}
\left(2 a_3 . \varepsilon \frac{\not a_3}{a_3.a_3} - \not \varepsilon
\right)\chi_4^{(k+1)}
\end{array}
\right.\cr
& \ \ \ \ 
\delta_\varepsilon \chi_l^{(3)}  = \varepsilon^n 
\Lambda^{-2/3} P_m  \chi_l^{(1)} 
\end{array}
\right\} l=0,1,2
\end{array}
\eeqa

\noindent
 However, it is interesting to note that if we 
choose the 3SUSY parameter $\varepsilon^m$ along the direction orthogonal to 
the hyperplane formed by $(a_0,a_1, a_2)$ in the Minkowski 4D space 
($\varepsilon=a_3$), then the 3SUSY transformations of the $\chi$'s read

\beqa
\begin{array}{ll}
\left.
\begin{array}{ll}
k=1,2 &\left\{
\begin{array}{l}
\delta_\varepsilon \chi_l^{(k)}  =
\Lambda^{1/3}\left(
{\not \varepsilon} \chi_l^{(k+1)} +2i
 {\not \varepsilon} \frac{{\not a_l}}{a_l^2} \chi_4^{(k+1)}\right) 
 \cr
\delta_\varepsilon \chi_4^{(k)}  = \Lambda^{1/3}{\not \varepsilon} 
\chi_4^{(k+1)}
\end{array}
\right.\cr
& \ \ \ \ 
\delta_\varepsilon \chi_l^{(3)}  = \varepsilon^n 
\Lambda^{-2/3} P_m  \chi_l^{(1)}.
\end{array}
\right\} l=0,1,2
\end{array}
\eeqa

\noindent
The usual 3SUSY transformations are thus retrieved in the limit of large
$a_l$'s.  This shows that at the level of the generating
functional for the correlation functions, one can treat the $a_l$'s 
as background fields. Formally, in the limit discussed above, their net effects
would be Dirac delta terms in the action leading to 3SUSY consistent 
constraints among the $\psi_l$ and $\chi^{(k)}_l$ fields. 

\noindent  
[Of course, the above ansatz, taken here for illustration, is not 
necessarily unique.]     



\section{Boson multiplets}
\renewcommand{\theequation}{4.\arabic{equation}}   
\setcounter{equation}{0} 
In the previous section, we were considering the fundamental representation
associated with the matrices (\ref{Mat6}) and (\ref{Mat6cc}), say
$\mathbf{\Psi}= \begin{pmatrix} \psi_1{}_\alpha \cr 
\bar \psi_2{}^{\dot \alpha} \cr \psi_3{}_\alpha \end{pmatrix}$ and
$\mathbf{\Psi^\prime}= \begin{pmatrix}\bar \psi^\prime_1{}^{\dot \alpha} \cr 
\psi^\prime_2{}_{ \alpha} \cr \bar \psi^\prime_3{}^{\dot \alpha} 
\end{pmatrix}$, {\it i.e.} with the vacuum $\Omega$ in the trivial 
representation
of the Lorentz algebra. In this section,  boson multiplets, will
be introduced, corresponding to a vacuum in the spinor representations
of the Lorentz algebra. This means that four types of boson
multiplets can be introduced: $\mathbf{\Psi} \otimes \Omega^\alpha,
\mathbf{\Psi^\prime} \otimes \Omega^\alpha$, with $\Omega^\alpha$ a left-handed
spinor and $\mathbf{\Psi} \otimes \bar \Omega_{\dot \alpha},
\mathbf{\Psi^\prime} \otimes \bar \Omega_{\dot \alpha}$  with 
$\bar \Omega_{\dot \alpha}$ a right-handed spinor.

For the multiplet associated with  
$\mathbf{\Psi}^\beta =\mathbf{\Psi} \otimes \Omega^\beta$,  we have
$\mathbf{\Psi}^\beta= \begin{pmatrix} \rho_1{}_{\alpha}{}^{ \beta} \cr 
 \bar \rho_2{}^{\dot \alpha \beta} \cr \rho_3{}_{\alpha }{}^{\beta} 
\end{pmatrix}$ and the transformation  under 3SUSY is 
$\delta_\varepsilon \mathbf{\Psi}^\beta = \varepsilon^m Q_m  \mathbf{\Psi}^\beta$
with $Q_m$ given in (\ref{Mat6}). This leads to

\beqa
\label{transfo2}
\delta_\varepsilon \rho_1{}_{\alpha}{}^{ \beta}& =& \Lambda^{1/3}
\varepsilon^m \sigma_m{}_{\alpha \dot \alpha}
\bar \rho_2{}^{\dot \alpha \beta} \nonumber \\
\delta_\varepsilon \bar \rho_2{}^{\dot \alpha \beta} &=&\Lambda^{1/3}
\varepsilon^m \bar \sigma_m{}^{\dot \alpha  \alpha} \rho_3{}_{\alpha}{}^{ \beta}
\\
\delta_\varepsilon \rho_3{}_{\alpha}{}^{ \beta}& =& \Lambda^{-2/3} \varepsilon^m P_m
\rho_1{}_{\alpha}{}^{ \beta} . \nonumber 
\eeqa

\noindent
Notice that $\rho_1, \bar \rho_2$ and $\rho_3$ are  not an irreducible representation
of $\mathfrak{sl}(2,\mathbb C) \cong \mathfrak{so}(1,3)$. We therefore
define 

\beqa
\rho_1&=& \varphi \,I_2 + \frac{1}{2} B_{mn} \,\sigma^{nm} \nonumber \\
\bar \rho_2&=& A^m \, \bar \sigma_m \\
\rho_3&=& \tilde \varphi \, I_2 + \frac{1}{2} \tilde B_{mn}\, \sigma^{nm}
 \nonumber
\eeqa

\noindent
with $I_2$ the two by two identity matrix,
$A^m$ a vector, $\varphi, \tilde \varphi$ two scalars and
$B_{mn},\tilde B_{mn}$ two self-dual two-forms.

Using the following dictionary to convert the spinor indices to vector
indices and {\it vice versa}, together with the trace properties of the
$\sigma$ matrices, we have 

\beqa 
\label{spin-vect}
\begin{array}{ll}
\varphi = \frac{1}{2} \mathrm{Tr}\left[\rho_1\right],&
\tilde \varphi = \frac{1}{2} \mathrm{Tr}\left[\rho_3\right],  \cr
B_{mn}=  \mathrm{Tr}\left[\sigma_{mn} \rho_1\right],&
\tilde B_{mn}=  \mathrm{Tr}\left[\sigma_{mn} \rho_3\right], \cr
A_m= \frac{1}{2} \mathrm{Tr}\left[\sigma_m \bar \rho_2\right],&
\end{array}
\eeqa

\noindent
with Tr the trace over the two by two matrices and spinors indices
contraction as in appendix B.
The tensor field $B_{mn}$ is self-dual (${}^\star B_{mn} \buildrel{\hbox{ def}} 
\over  = \frac{1}{2} \varepsilon_{mnpq} B^{pq}= i B_{mn}$) because of 
the property of
self-duality of $\sigma_{mn}$, see the appendix B .

Using the correspondence (\ref{spin-vect}),  the
transformations (\ref{transfo2}) become

\beqa
\label{transfo2vect}
\delta_\varepsilon \varphi &=&\Lambda^{1/3}  \varepsilon^m A_m \nonumber \\
\delta_\varepsilon B_{mn} &=& - \Lambda^{1/3}\left(
\varepsilon_m A_n - \varepsilon_n A_m \right) +\Lambda^{1/3} i  
\varepsilon_{mnpq} \varepsilon^p A^q  \nonumber \\
\delta_\varepsilon A_m &=& \Lambda^{1/3}\left( \varepsilon_m \tilde \varphi +
 \varepsilon^n \tilde B_{mn} \right)   \\
\delta_\varepsilon \tilde \varphi &=& \Lambda^{-2/3} 
\varepsilon^m P_m \varphi \nonumber \\
\delta_\varepsilon  \tilde B_{mn} &=& \Lambda^{-2/3}\varepsilon^p P_p B_{mn}
\nonumber 
\eeqa

In a similar way, we consider a multiplet (CPT conjugate of the previous)
constructed from   
$\mathbf{\Psi}_{\dot \beta} =\mathbf{\Psi^\prime} 
\otimes \Omega_{\dot \beta}= \begin{pmatrix} \bar 
\rho_1{}^{\dot\alpha}{}_{ \dot \beta} \cr 
  \rho_2{}_{ \alpha \dot  \beta} \cr \bar 
\rho_3{}^{\dot \alpha }{}_{\dot \beta} 
\end{pmatrix}$. As before, we introduce the following fields

\beqa
\begin{array}{ll}
\bar \rho_1= \varphi^\prime \,\bar I_2 + \frac{1}{2} B^\prime_{mn} \,
\bar \sigma^{nm}& \ \ \ \ \ \ \ \ \ \ \ \ \ \cr
\rho_2= A^m \,  \sigma_m \cr
\bar \rho_3= \tilde \varphi \, \bar I_2 + \frac{1}{2} \tilde B_{mn}\, 
\bar \sigma^{nm}
 \end{array}
\nonumber \\
\\
\begin{array}{ll}
\varphi^\prime  = \frac{1}{2} \mathrm{Tr}\left[\bar \rho_1\right],&
\tilde \varphi^\prime = \frac{1}{2} \mathrm{Tr}\left[\bar \rho_3\right],  \cr
B^\prime_{mn}=  \mathrm{Tr}\left[\bar \sigma_{mn} \bar \rho_1\right],&
\tilde B^\prime_{mn}=  
\mathrm{Tr}\left[\bar \sigma_{mn} \bar \rho_3\right], \cr
A^\prime_m= \frac{1}{2} \mathrm{Tr}\left[\bar \sigma_m \rho_2\right].&
\end{array} \nonumber 
\eeqa

\noindent
Their transformations are found to be 

\beqa
\label{transfo2vect2}
\delta_\varepsilon \varphi^\prime &=&  \Lambda^{1/3}
\varepsilon^m A^\prime_m \nonumber \\
\delta_\varepsilon B^\prime_{mn} &=& - \Lambda^{1/3}\left(
\varepsilon_m A^\prime_n - \varepsilon_n A^\prime_m \right) -\Lambda^{1/3} i  
\varepsilon_{mnpq} \varepsilon^p A^\prime{}^q  \nonumber \\
\delta_\varepsilon A^\prime_m &=&  \Lambda^{1/3}\left(
\varepsilon_m \tilde \varphi^\prime + 
\varepsilon^n \tilde B^\prime_{mn}\right)   \\
\delta_\varepsilon \tilde \varphi^\prime &=& \Lambda^{-2/3}\varepsilon^m 
P_m \varphi^\prime \nonumber \\
\delta_\varepsilon  \tilde B^\prime_{mn} &=& \Lambda^{-2/3}\varepsilon^p P_p 
B^\prime_{mn}.
\nonumber 
\eeqa

\noindent
The fields  $B^\prime, \tilde B^\prime$ are in this case
anti-self-dual.\\
The second bosonic multiplet being
CPT conjugate to the first one, we have
$\left(\bar \rho_1{}^{\dot \alpha}{}_{\dot \beta}\right)^\star=
\rho_1{}^{\alpha}{}_{\beta},
\left(\rho_{2 \alpha \dot \beta}\right)^\star=
\bar \rho_{2 \dot \alpha  \beta}$ and
$\left(\bar \rho_3{}^{\dot \alpha}{}_{\dot \beta}\right)^\star=
\rho_3{}^{\alpha}{}_{\beta}$. 
[$B^\star$, the complex conjugate of $B$,
is  not to be confused with ${}^\star B$, the dual of $B$.]
This means, paying attention to the position
of the indices and the conventions given in appendix B, that we have

\beqa
\label{cc}
\begin{array}{ll}
\varphi^\prime{}^\star = - \varphi,&
 \tilde \varphi^\prime{}^\star = -\tilde \varphi, \cr
B_{mn}^{\prime \, \star}= -B_{mn},&
\tilde B_{mn}^{\prime \, \star}= -\tilde B_{mn}, \cr
A^{\prime \, \star}_m= A_m.
\end{array}
\eeqa

\noindent
These relations are compatible with the transformations laws given
in (\ref{transfo2vect}) and (\ref{transfo2vect2}) since
$\varepsilon_n{}^\star = - \varepsilon_n$ and $P_n= -i \frac{\partial}{\partial
x^n}$.

The two-forms which appear in the bosonic multiplet can be treated
either as matter fields \cite{lemes} or as gauge fields in the context
of extended objects \cite{nambu}. Here we will consider only the latter
interpretation since the two-forms sit in the same
multiplet as the $A$'s which we attempt to identify with the  usual gauge fields.
However, we will not consider in this paper the coupling to any matter
fields, whether point-like or not.  
Notice that the gauge transformation $B_{mn} \to B_{mn} + \partial_m \chi_n - 
\partial_n \chi_m$ which  preserves the associated field strength $H$
 (see below) does not preserve the self-duality of $B$. 
Conversely, a self-duality preserving transformation ($B_{mn} \to B_{mn} + 
(\partial_m \chi_n -\partial_n \chi_m )-
i \varepsilon_{mnpq}\partial^p \chi^q$) does not preserve the field strength.
(Similar remarks hold for $B^\prime$.)

To identify the $A$'s with some gauge fields 
we introduce the real-valued fields 

\beqa
A_-=i\frac{A-A^\prime}{\sqrt{2}} &,&
A_+= \frac{A+A^\prime}{\sqrt{2}}, \nonumber \\
B_-= \frac{B-B^\prime}{\sqrt{2}} &,& 
B_+= i\frac{B+B^\prime}{\sqrt{2}}, \nonumber \\
\tilde B_-=  \frac{\tilde B-\tilde B^\prime}{\sqrt{2}} &,& 
\tilde B_+=  i\frac{\tilde B+\tilde B^\prime}{\sqrt{2}}, \\
\varphi_-= \frac{\varphi-\varphi^\prime}{\sqrt{2}} &,& 
\varphi_+= i\frac{\varphi+\varphi^\prime}{\sqrt{2}}, \nonumber \\
\tilde \varphi_-=  \frac{\tilde \varphi-\tilde \varphi^\prime}{\sqrt{2}} &,& 
\tilde \varphi_+=  i\frac{\tilde \varphi+\tilde \varphi^\prime}{\sqrt{2}}. \nonumber
\eeqa 


\noindent
These new fields form now one (reducible) multiplet of 3SUSY,  
 with ${}^\star B_- =  B_+$.
The corresponding  two- and three-form field strengths read

\beqa
\label{curl}
&&F_{\pm}{}_{mn}=\partial_m A_{\pm}{}_n -\partial_n A_{\pm}{}_m, \  
\nonumber \\
&&H_{\pm}{}_{mnp}= \partial_m B_{\pm}{}_{np} + \partial_{n} 
B_{\pm}{}_{pm} + 
\partial_p B_{\pm}{}_{mn}.
\eeqa 

\noindent
They  are  invariant under the gauge transformations

\beqa
\label{form}
\varphi_\pm &\to & \varphi_\pm   \nonumber \\
A_{\pm}{}_m &\to & A_{\pm}{}_m + \partial_m \chi_{\pm} \\ 
B_{\pm}{}_{mn} &\to & B_{\pm}{}_{mn} + 
(\partial_m \chi_{\pm \,n} -\partial_n \chi_{\pm \, _m} ) \nonumber
\eeqa

\noindent
where $\chi_{\pm}$ ($\chi_{\pm}^m$) are arbitrary scalar (vector)
functions  ($ \chi^m_{-}$ and    $\chi^m_+$ can still be related in order
to preserve the duality relations between $B_-$ and $B_+$)\footnote{
Note that the gauge transformations (\ref{form}) correspond naturally
to the zero-, one- and two-form character of the components of the 3SUSY
gauge multiplet.}.
 
In a similar way we introduce the field strength
$\tilde H_-{}_{mnp}, \tilde H_+{}_{mnp}$, as well as the dual fields 
${}^\star H_-{}_m,
{}^\star H_+{}_m,
{}^\star\tilde  H_-{}_m, {}^\star \tilde H_+{}_m$
(where ${}^\star H_m \equiv \frac{1}{6} \varepsilon_{mnpq} H{}^{npq}$.
For instance ${}^\star \tilde H_-{}_m = i \partial^n B_+{}_{m n}$).
We consider now the following local gauge invariant and zero graded 
Lagrangian,

\beqa
\label{Lbos}
{\cal L}& = &\partial_m \varphi_- \partial^m \tilde \varphi_-  -
           \partial_m \varphi_+  \partial^m \tilde \varphi_+ 
+\frac{1}{4} F_{-}{}_{mn}F_-{}^{mn} -\frac{1}{4} F_{+}{}_{mn}F_+{}^{mn} \\
&+& \frac{1}{12} H_-{}_{mnp} \tilde H_-{}^{mnp} 
-\frac{1}{12} H_+{}_{mnp} \tilde H_+{}^{mnp}
-\frac{1}{2} {}^\star H_-{}_m  {}^\star \tilde H_-{}^{m}+
\frac{1}{2} {}^\star H_+{}_m  {}^\star \tilde H_+{}^{m} \nonumber
\eeqa

\noindent

After some algebra, one obtains the 3SUSY variation of the Lagrangian,
up to a surface term,

\beqa
\label{fusy-jauge}
\delta_\varepsilon {\cal L} = (\delta_\varepsilon \partial_m A_-{}^n)  
\partial{}_n A_-{}^m - (\delta_\varepsilon \partial_m A_+{}^n) 
\partial_n A_+{}^m
\eeqa

\noindent
meaning that the gauge-invariant Lagrangian (\ref{Lbos}) is not 3SUSY
invariant. We checked that this result remains unavoidable even if
we added the two remaining boson multiplets 
${\bf \Psi}^\prime \otimes \Omega^\beta$ and 
${\bf \Psi} \otimes \Omega_{\dot\beta}$.\\

It is now interesting to note, that up to surface terms,

\beqa
\label{fusy-jauge2}
\delta_\varepsilon {\cal L} =  
\frac{1}{2}\delta_\varepsilon (\partial_n A_-{}^n)^2 -
\frac{1}{2} \delta_\varepsilon (\partial_n A_+{}^n)^2 
\eeqa

\noindent
This seems to indicate that 3SUSY invariance requires the usual
't Hooft-Feynman gauge fixing term:

\beqa
\label{tf}
{\cal L}_{\mathrm{fsusy}} =  {\cal L} + {\cal L}_{\mathrm{g.f}}(\xi=1), 
\eeqa

\noindent where  ${\cal L}$  is defined in (\ref{Lbos}) and 

\beqa
 {\cal L}_{\mathrm{g.f}}(\xi)=
-\frac{1}{2 \xi } (\partial_n A_-{}^n)^2  +
                             \frac{1}{2 \xi } (\partial_n A_+{}^n)^2
\eeqa


\noindent
leading to 
$$\delta_\varepsilon {\cal L}_{\mathrm{fsusy}} = 0 \mathrm{~(up~to~a~ 
surface~term.)}$$ 

\noindent
We do not dwell further here on this intriguing result, that is 
{\sl one symmetry (3SUSY) leads naturally to a physically necessary gauge 
fixing of another (namely $U(1)$ associated with the vector field $A$)} and 
thus 
eliminating the intrinsic unphysical redundancies endemic to gauge 
theories\footnote{For comments on interesting proposals to tackle
 the question of gauge redundancies, in the context of standard SUSY, see for 
instance \cite{polchinski}, section 4.4. 
It is also worth noting that, even in the absence of (super)symmetries,
such gauge fixing conditions appeared naturally in the action-at-a-distance 
formulation of coupling of a  point-like particle (resp. a string) to a 
vector (resp. a two-form). In particular, in the case of the vector field this 
leads to the Fermi Lagrangian
(which is precisely  $-\frac14 F_{mn}F^{mn}
+ {\cal L}_{\mathrm{g.f}}(\xi=1)$, up to a surface-term), and in the 
case of the two-form to a
 Lagrangian 
which is analogous to $\frac{1}{12} H_+{}_{mnp} \tilde H_+{}^{mnp}-
\frac{1}{2} {}^\star H_-{}_m  {}^\star \tilde H_-{}^{m}$, 
see Kalb and Ramond in \cite{nambu}.}. 

\noindent
One should note, though, the relative minus signs in front of the kinetic
terms of the vector fields in (\ref{Lbos}) which endanger {\sl a priori} 
the boundedness from below of the energy density of the 
``electromagnetic" fields. This difficulty does not have a clear physical
interpretation as long as interaction terms have not been included, and 
necessitates a more careful study of the field manifold
associated with the density energy.  However, it may suggest that some field
combinations are dynamically driven to very large values and should be
reinterpreted as decoupling from the physical system.\\

Before concluding this section, we point out that one can partially couple
in a 3SUSY invariant way the two-forms associated with the two CPT conjugate
multiplets. Namely, defining
\beqa
{\cal L}_B' &=&   H{}_{mnp} \tilde H^\prime{}^{mnp}
+  H^\prime{}_{mnp} \tilde H{}^{mnp}, \nonumber \\
{}^\star {\cal L}_B' &=&   {}^\star H{}_{mnp} {}^\star \tilde H^\prime{}^{mnp}
+  {}^\star H^\prime{}_{mnp} {}^\star \tilde H{}^{mnp}, \nonumber
\eeqa

\noindent
one finds that the Lagrangian

\beqa
{\cal L}' =  \frac{1}{12} {\cal L}_B' 
- \frac12 {}^\star {\cal L}_B'
\eeqa

\noindent
is by itself gauge and 3SUSY invariant. 


Including similar couplings in
the $\varphi$ and $A$ sectors, it is possible to find combinations  with the 
Lagrangian in (\ref{Lbos}) where the vector field kinetic terms have the
correct signs. However, the so obtained Lagrangian is no more 3SUSY invariant
even including gauge fixing terms.

\section{A comment on Noether currents}
\renewcommand{\theequation}{5.\arabic{equation}}
\setcounter{equation}{0}

The 3SUSY algebra we have studied  has one main difference 
with the usual Lie (super)algebra: it does not close through quadratic,
but rather cubic, relations. Moreover, it might be possible that
some usual results of Lie (super)algebra do not apply straightforwardly.
One example is the  Noether currents and their associated algebra.
Indeed, according to 
Noether theorem, to all the symmetries correspond conserved currents.
The symmetries are then  generated by charges which are expressed in 
terms of the fields.
By the spin-statistics theorem,  
fields having integer spin  close with commutators whilst fields
with half-integer spin under anticommutators. So it seems that the current
algebra automatically leads to Lie (super)algebras.
In our case, however, things might look different since we are not dealing 
with a Lie algebra and yet we have only bosonic operators.
It is thus interesting to understand how this unusual feature
translates. 

We will actually start from the classical field theory case,
then go to the quantum case through the usual canonical
quantization procedure.
Let us consider a  general Lagrangian ${\cal L}$,
at the classical level,  and  construct through the standard procedure 
the Noether charges $\hat{Q}_A$ 

\begin{equation}
\label{charge}
\varepsilon^A \hat{Q}_A = 
\int d^3x \frac{\delta {\cal L}}{\delta\partial_0 \Psi_i} 
\delta_\varepsilon \Psi_i 
\end{equation}

\noindent
associated with the symmetry transformations

\begin{equation}
\delta_\varepsilon \Psi_i = \varepsilon^A Q_A 
\Psi_i  \label{trans1}
\end{equation}

\noindent
Here $Q_A$ generate the symmetry algebra in some
appropriate (matrix) representation and $\Psi_i$
is a generic field.
In particular, $Q$ could be associated
with the 3SUSY symmetry transformations.
Upon use of eq.(\ref{trans1}) in eq.(\ref{charge}),   
one gets

\begin{equation}
\label{charge1}
\hat{Q}_A = 
\int d^3x \Pi_i(x)  Q_A \Psi_i(x) 
\end{equation}

\noindent
where $\Pi_i(x) \equiv \frac{\delta {\cal L}}{\delta\partial_0 \Psi_i}$
is the conjugate momentum. Equation (\ref{charge1}) is the general
relation between $\hat{Q}_A$ and $Q_A$.
 Now from eqs.(\ref{trans1}, \ref{charge1}) one readily  gets

\begin{equation}
\left\{\varepsilon^B \hat{Q}_B, \Psi_i(x)\right\}_{\mbox{p.b}} = 
\delta_\varepsilon \Psi_i = \varepsilon^A Q_A 
\Psi_i \label{trans2}
\end{equation}

\noindent
meaning that {\sl at the classical level} the Noether charges $\hat{Q}_B$
generate the considered algebra through Poisson brackets.
The quantum case is then obtained in the usual canonical way by replacing
the Poisson bracket by the commutator $[,]$ in eqs.(\ref{trans2}) 
where now $\hat Q, \hat P$ and $\Psi_i$ operate on some Fock 
space, leading to 
\begin{equation}
[\varepsilon^A \hat{Q}_A, \Psi_i(x)] = 
\delta_\varepsilon \Psi_i \label{trans3}
\end{equation}

\noindent
Before going further in the 3SUSY case, let us make a quick digression 
and recall some features
related to the (more conventional) Lie (super)algebra case.
If $Q_a$ and $\hat Q_a$ were associated with Lie (super)algebra with
structure constants  $f_{ab}{}^c$, the algebra
would have been realized as follows

\begin{equation}
\label{dLie}
 [\delta_a, \delta_b] \Psi_i \equiv 
\delta_a (\delta_b \Psi_i) -  \delta_b (\delta_a \Psi_i) =
 f_{a b}{}^{ c} \delta_c \Psi_i 
\end{equation}

\noindent
where $\Psi_i$ is now   a field operator and $\delta$ is 
given by (\ref{trans3}).
\noindent
Equivalently, equation (\ref{dLie}) reads then  

\begin{equation}
\label{Lie}
[\hat Q_a, [\hat Q_b, \Psi_i]] - [\hat Q_b, [\hat Q_a, \Psi_i]] =
     f_{a b}{}^{ c} [\hat Q_c, \Psi_i] 
\end{equation}

\noindent
Now due to Jacobi identity ($ [A, [B, C]] + \mbox{cyclic} = 0$), one can recast
the left-hand side of Eq.(\ref{Lie}) in the form 
$ [[\hat Q_a, \hat Q_b], \Psi_i]$
to get

\begin{equation}
[[\hat Q_a, \hat Q_b], \Psi_i] = f_{a b}{}^{ c} [\hat Q_c, \Psi_i] 
\; \; \; \; \; \; \; \;  \mbox{(for any $\Psi_i$)} \label{eq4}
\end{equation}

\noindent
This means that 

\begin{equation}
[\hat Q_a, \hat Q_b] = f_{a b}{}^{ c} \hat Q_c  \label{eq5}
\end{equation}

\noindent
at least on some (sub-)space of field operators $\Psi_i$. \\

For the 3SUSY algebra we are considering, the analogy with the steps 
described above stops at Eq.(\ref{Lie}). Although we do have 
(generalized) Jacobi identities (see Eq. (\ref{eq:J})), these do not 
help in going from equations

\beqa
\label{F-com}
\left( \delta_m. \delta_n.
\delta_r + \mathrm{perm} \right) \mathbf{ \Psi}& =& \\
\left[\hat Q_m, \left[\hat Q_n, \left[\hat Q_r,  
\mathbf{ \Psi} \right]\right]\right]+ \mathrm{perm}& =&
\eta_{m n} \left[\hat P_r,\mathbf{ \Psi} \right] +  
\eta_{m r} \left[\hat P_n,\mathbf{ \Psi} \right] + 
\eta_{r n} \left[\hat P_m, \mathbf{ \Psi} \right] \nonumber
\eeqa

\noindent
which is the  analog of 
Eq.(\ref{Lie})
(with $\hat P_m$ the generators of the Poincar\'e translations),
 to an equation of the form

\begin{equation}
[\left\{\hat Q_m, \hat Q_n,\hat Q_r\right\}, \Phi]
=[\eta_{m n} \hat P_r +  \eta_{m r} \hat P_n + 
\eta_{r n} \hat P_m, \Phi] 
\label{eq6}
\end{equation}

\noindent
which would have been the analog of Eq.(\ref{eq4}). 
In other words, 
the quantized version of the Noether charges algebra is just (\ref{F-com})
and cannot be cast simply in a $\mathbf{ \Psi}$ independent
 form\footnote{Note that this aspect is not due to the quantization procedure.
It is already present at the classical level.
In this case the Noether charges algebra

\begin{equation}
\left\{\hat Q_m, \left\{\hat Q_n, \left\{\hat Q_r, 
\Psi_i \right\}_{\mbox{p.b}}\right\}_{\mbox{p.b}}
\right\}_{\mbox{p.b}} +  perm.
= \left\{ \eta_{m n} \hat P_r +  \eta_{m r}  \hat P_n + 
\eta_{r n}  \hat P_m, \Psi_i \right\}_{\mbox{p.b}}
\nonumber
\end{equation}
\noindent
cannot be recast in the form of the third equation of (\ref{F-poin})
simply because we do not have a notion of Poisson bracket with three
entries.  (Such a notion would require for instance a bi-Hamiltonian formalism
\cite{nb}, which is not the track we follow here.)
}. We should stress at this level that the difference with the conventional 
algebras we are pointing out, does not mean the absence of a realization of the
3SUSY algebra in terms of Noether charges.
Indeed, starting with the abstract algebra (\ref{F-poin}), we
can represent it by some matrices as in section 2 (see {\it e.g.}
(\ref{Mat12})). In this case the product of two 
transformations will be given
by $\delta_n \delta_m \Psi = Q_n Q_m \Psi$
and the algebra will be realized as in  (\ref{ext-susy}). 
In the case of Noether charges, the product of two transformations 
is given by 
$\delta_n \delta_m \Psi = \left[\hat Q_n, \left[\hat Q_m, \Psi \right]\right]$
(see (\ref{trans3})), leading to the realization (\ref{F-com}) of the 
algebra (\ref{ext-susy}).
[We explicitly checked (\ref{F-com}) on particular Lagrangians such
as (\ref{Lbos}) using the usual canonical commutators.]\\

\noindent
The point which is potentially tricky is the fact that (\ref{F-com}) cannot
be made $\mathbf{ \Psi}$ independent in general. Thus the construction of
3SUSY representation states in the Fock space requires some care.
Actually (\ref{F-com}) becomes equivalent to (\ref{ext-susy}) (with
$ Q \to \hat Q, P \to \hat P$), when acting on {\sl one particle} Fock states.
Thus one can construct {\sl one particle} state representations using
a  $\mathbf{ \Psi}$ independent form of the algebra. Now the difference with
the conventional algebras is that N-particle state representations will
not be trivially obtained from the usual tensor product of one-particle
states (a difference which holds even at the classical level). 
However, one can show that  when acting on N-particle states, the 
deviation from (\ref{ext-susy}) (with $ Q \to \hat Q, P \to \hat P$) is 
expressed in terms of the action of the algebra on (N-1)-particle states. The
construction is thus obtained iteratively. We will not detail further here
these issues which are outside of the scope of the present paper.

\noindent

\section{Conclusion}

In the present paper we have constructed the first four dimensional
examples of field theories which are invariant under a non trivial extension
of the Poincar\'e algebra that is {\sl not} the supersymmetric one.
In particular, we constructed representations in the form of 
fermion and boson multiplets and their associated invariant actions.
In the fermionic case we identified potential difficulties:
for the simplest invariant action, the fermionic fields behave like
anticommuting spinless fields {\it i.e.} ghosts.
This could be resolved through couplings to background fields.
For the bosonic multiplet, we were able to construct a $U(1)$ gauge 
invariant action compatible with our symmetry provided that a 't Hooft-Feynman
gauge fixing term is added. The natural appearance of 2-form fields could
be a hint that both point particle and one-dimensional extended matter objects
should be included in the theory. 
Strictly speaking, the existence proof of such theories is not complete
until a model with interaction terms has been constructed.

\paragraph*{Acknowledgements}  
G. Mennessier, A. Neveu and P. Townsend are gratefully acknowledge
for useful remarks and comments.

\appendix
\section{Algebraic foundation  of FSUSY}
\renewcommand{\theequation}{A.\arabic{equation}}
We summarize here for completeness the salient algebraic features of
FSUSY and its underlying $F-$Lie algebra structures.
More details can be found in \cite{flie1, flie2}.
\subsection{The algebra}
The general definition of  $F-$Lie algebras, the abstract algebraic
structure underlying FSUSY, was given in \cite{flie1, flie2} together with
an inductive way to construct $F-$Lie algebras associated with {\it any}
Lie algebra or Lie superalgebra. We do not
want  to go into the  detailed definition of this structure here
and will  only recall the basic points, useful for the sequel.
More details can be found in~\cite{flie1, flie2}. 
We consider $F$  a positive integer and define 
$q=e^{\frac{ 2 \pi i}{F}}$. The algebra $\mathfrak{g}= \mathfrak{g}_0 \oplus
\mathfrak{g}_1$ is called an $F-$Lie algebra if the following four
properties hold:

\begin{enumerate}
\item $\mathfrak{g}_0$ is a Lie algebra;
\item $\mathfrak{g}_1$ is a representation of $\mathfrak{g}_0$;
\item  There exists  a multilinear,  $\mathfrak{g}_0-$equivariant  $F-$fold
({\it i.e.} which respect
the action  of  $\mathfrak{g}_0$) map
 $\left\{ \cdots \right\}: {\cal S}^F\left(\mathfrak{g}_1\right)
\rightarrow \mathfrak{g}_0$ from
${\cal S}^F\left(\mathfrak{g}_1\right)$ into  $\mathfrak{g}_0$.
In other words, we  assume that some of the elements of the Lie algebra
$\mathfrak{g}_0$ can be expressed as $F-$th order symmetric products of
``more fundamental generators''. Here  ${ \cal S}^F(\mathfrak{g}_1)$ denotes
the $F-$fold symmetric product of $\mathfrak{g}_1$. It can be  easily seen 
that this 
bracket simply corresponds to the anticommutator, so to Lie superalgebras, when
$F=2$.
\item  For $b_i \in \mathfrak{g}_0$ and $a_j \in \mathfrak{g}_1$ the following
``Jacobi identities'' hold:
 
\beqa
\label{eq:J}
&&\left[\left[b_1,b_2\right],b_3\right] +
\left[\left[b_2,b_3\right],b_1\right] +
\left[\left[b_3,b_1\right],b_2\right] =0 \nonumber \\
&&\left[\left[b_1,b_2\right],a_3\right] +
\left[\left[b_2,a_3\right],b_1\right] +
\left[\left[a_3,b_1\right],b_2\right]  =0 \nonumber \\
&&\left[b,\left\{a_1,\dots,a_F\right\}\right] =
\left\{\left[b,a_1 \right],\dots,a_F\right\}  +
\dots +
\left\{a_1,\dots,\left[b,a_F\right] \right\} \\
&&\sum\limits_{i=1}^{F+1} \left[ a_i,\left\{a_1,\dots,
a_{i-1},
a_{i+1},\dots,a_{F+1}\right\} \right] =0. \nonumber
\eeqa

\end{enumerate}

\noindent
It can be seen that $F-$Lie algebras admit a $\mathbb Z_F$ grading:
$\mathfrak{g}_0$ being of grade $0$ and $\mathfrak{g}_1$ of grade
$1$.  This means that there exists a grading map $g$ such that
$g a g^{-1}=q a , a \in \mathfrak{g}_1$ and
$g b g^{-1}= b, b \in \mathfrak{g}_0$, with $q$ a 
primitive  root of unity ($q^F=1$).
This notion can also be introduced in a more formal way, see
\cite{flie1, flie2} for more details. From now on, $\mathfrak{g}_0$ will 
be called the bosonic sector of $\mathfrak{g}$ and
$\mathfrak{g}_1$ the graded sector \footnote{In general, an $F-$Lie algebra admits the decomposition
$\mathfrak{g}=\mathfrak{g}_0 \oplus \mathfrak{g}_1 \oplus \cdots \oplus
  \mathfrak{g}_{F-1}$
see \cite{flie1}, \cite{flie2}}.  \\

Having defined the  structure of $F-$Lie algebras, we could construct
in a systematic way, explicit examples of $F-$Lie algebras associated with Lie
(super)algebras \cite{flie2}. Then, among these families of 
finite-dimensional examples one can 
identify $F-$Lie algebras that could generate a
non-trivial extension of the Poincar\'e algebra. Two such examples were
given in \cite{flie2}. Here, we just consider one of these examples, where
$\mathfrak{g}_0=\mathfrak{sp}(4,\mathbb R)$ and 
$\mathfrak{g}_1=\mathrm {ad} \left(\mathfrak{sp}(4,\mathbb R) \right)$, 
with ad, the
adjoint representation of $\mathfrak{sp}(4,\mathbb R)$. If we denote
$J_a, \ a=1, \cdots, 10$ a basis of $\mathfrak{sp}(4,\mathbb R)$, 
$A_a$ the corresponding basis for $\mathfrak{g}_1$, and 
$g_{ab}= {\mathrm Tr} \left(A_a A_b\right)$ the Killing form of 
$\mathfrak{sp}(4,\mathbb R)$, then  the $F-$Lie algebra of order $3$ 
$\mathfrak{g}= \mathfrak{g}_0 \oplus \mathfrak{g}_1$ reads

\beqa
\label{flie_sp}
\left[J_a, J_b \right] = f_{ab}^{\ \ \ c} J_c, \qquad
 \left[J_a, A_b \right] = f_{ab}^{\ \ \ c} A_c, \qquad
\left\{A_a, A_b, A_c\right\}= g_{ab}J_c + g_{ac} J_b + g_{bc} J_a
\eeqa

\noindent
where $f_{ab}^{ \ \ c}$ are the structure constant of $\mathfrak{g}_0$.

Observing that $\mathfrak{so}(1,3) \subset \mathfrak{so}(2,3) 
\cong \mathfrak{sp}(4)$, and that the $(1+3)D$ Poincar\'e algebra is related 
to $\mathfrak{sp}(4)$ through an  In\"on\"u-Wigner contraction, one can expect 
to obtain, from the $F-$Lie algebra (\ref{flie_sp}),
 an extension of the Poincar\'e algebra \footnote{In the same way
$N-$extended supersymmetric algebra can be obtained though an In\"on\"u-Wigner
contraction of the superalgebra $\mathfrak{osp}(N|4)$.}.
Using vector indices of $\mathfrak{so}(1,3)$ coming from the inclusion
$\mathfrak{so}(1,3) \subset \mathfrak{so}(2,3) 
\cong \mathfrak{sp}(4)$, the bosonic part of
$\mathfrak{g}$ is generated by $J_{m n}, J_{m 4}$, with  
$m, n =0,1,2,3$ and the graded part by $A_{m n}, A_{4 m}$ 
 ($J_{mn}=-J_{nm}$ and $A_{mn}=-A_{nm}$).
Letting $\lambda \to 0$
after  the In\"on\"u-Wigner contraction,

\beqa
\begin{array}{ll}
J_{m n} \to L_{m n},& J_{m 4} \to \frac{1}{\lambda} P_m 
\\
A_{m n} \to \frac{1}{\sqrt[3]{\lambda}} Q_{m n},& 
A_{4 m} \to \frac{1}{\sqrt[3]{\lambda}} Q_{m},
\end{array}
\eeqa

\noindent   
one sees that  
$L_{m n }$ and $P_m$ generate the $(1+3)D$ Poincar\'e
algebra and that $Q_{m n}, Q_m$ are the fractional supercharges
 in  respectively  the adjoint and vector representations of
$\mathfrak{so}(1,3)$.
This $F-$Lie algebra of order three is therefore a non-trivial
extension of the Poincar\'e algebra where translations are cubes
of more fundamental generators. The subspace  generated by 
$L_{m n}, P_m, Q_m$ is also an $F-$Lie algebra of order three
extending the Poincar\'e algebra  in which
the trilinear symmetric brackets have the simple form: 

\begin{eqnarray}
\label{F-poin2}
&&\left[L_{mn}, L_{pq}\right]=
\eta_{nq} L_{pm}-\eta_{mq} L_{pn} + \eta_{np}L_{mq}-\eta_{mp} L_{nq},
\ \left[L_{mn}, P_p \right]= \eta_{np} P_m -\eta_{mp} P_n, \nonumber \\
&&\left[L_{mn}, Q_p \right]= \eta_{np} Q_m -\eta_{mp} Q_n, \ \
\left[P_{m}, Q_n \right]= 0, \\
&&\left\{Q_m, Q_n, Q_r \right \}=
\eta_{m n} P_r +  \eta_{m r} P_n + \eta_{r n} P_m, \nonumber
\end{eqnarray}

\noindent
where $\eta_{m n}$ is  the Minkowski metric.
We should mention that this algebra can  also be considered in {\it any}
space-time dimensions. {}For the purpose of this paper, we
consider only the
$(1+3)$ dimensional case. 

\subsection{Representations}
The next step in the construction of  an invariant action under 
3SUSY transformations, is to construct the representations of
the 3SUSY algebra (\ref{F-poin2}).
A representation of an $F-$Lie algebra 
 $\mathfrak{g}$  is  a linear map
$\rho : ~ \mathfrak{g} \to \mathrm{End}(H)$ ($ \mathrm{End}(H)$ being
the space of linear applications of $H$ into $H$), 
and an  
automorphism $\hat g$ such that $ \hat g^F=1$. 
It satisfies 

\beqa      
\label{eq:rep}
\begin{array}{ll}
\mathrm{a)}& \rho\left(\left[x,y\right]\right)= \rho(x) \rho(y)-  
\rho(y)\rho(x) \cr
\mathrm{b)}& \rho \left\{a_1.\cdots,a_F\right\}=
 \sum \limits_{\sigma \in S_F}
\rho\left(a_{\sigma(1)}\right) \cdots \rho\left(a_{\sigma(F)}\right) \cr
\mathrm{d)}& \hat g \rho\left(s\right) \hat g^{-1} =
\rho\left(g\left(s\right)\right)
\end{array}
\eeqa

\noindent
($S_F$ being the group of permutations of $F$ elements). 
As  a consequence,  
since the eigenvalues of $\hat g$ are $\mathrm{F}^{\mathrm{th}}-$
roots of unity, we have  the following decomposition 

$$H= \bigoplus \limits_{k=0}^{F-1} H_k,$$

\noindent
where $H_k=\left\{\left|h\right> \in H ~:~ 
\hat g\left|h\right>=q^k \left|h\right> \right\}$.
The operator $N \in \mathrm{End}(H)$
defined by $N\left|h\right>=k
\left| h \right>$ if $\left|h\right> \in H_k$
is the  ``number operator'' (obviously $q^N=\hat g$).
Since $\hat g \rho(b)= \rho(b) \hat g, \forall b \in \mathfrak{g}_0$
each $H_k$ provides a representation of the Lie algebra $\mathfrak{g}_0$. 
Furthermore, for $a \in \mathfrak{g}_1$,
 $\hat g \rho(a)=q \rho(a) \hat g$ and so
we have 
$\rho(a) .H_k\ \subseteq 
H_{k+1 ({\mathrm{mod~} F)}}$  \\

Before constructing  representations of the algebra  (\ref{F-poin2}),
some general comments are in order. Firstly, observing that the operator
$P^2$ is a Casimir operator, all states in an irreducible representation
have the same mass. Secondly, writing $g=e^{i \frac{2 \pi}{3} N}$, with $N$
the number operator, and using the cyclicity of the trace it is easy to prove
that $\mathrm{Tr}(g)=0$ \cite{fsusy3d}. Thus
all the $H_i$ have the same dimension. Here we assume that 
we have a finite dimensional representation in order not to have problems with
the trace of infinite dimensional matrices (see below).

To obtain  representations of the algebra  (\ref{F-poin2}), we rewrite
the RHS of the trilinear bracket as  
$\left\{Q_m, Q_n, Q_r \right \}=f_{mnr} = f_{mnr}^{\ \ \ \ s}P_s$,
with  
$f_{mnr}{}^{ s}= \eta_{m n} \delta_r{}^{ s} +  \eta_{m r} \delta_n{}^{ s} + 
\eta_{r n} \delta_m{}^{ s}$.  This substitution shows that to the 
symmetric tensor $f_{mnr}$ is associated  the cubic polynomial 
$f(v^0,v^1,v^2,v^3)= f_{mnr} v^m v^n v^r = 3 (v.P)(v.v)$. Moreover, the
algebra (\ref{F-poin2}) simply  means that $f(v)= \left(v^m Q_m \right)^3$,
as can be verified by  developing the cube and identifying all terms,
using the trilinear bracket.
The generators  $Q_m, m=0, \cdots, 3$ which 
are associated with the  variables $v^m,m=0, \cdots, 3$, then generate
an extension of the Clifford algebra  
called Clifford algebra of the polynomial $f$ 
(denoted ${\cal C}_f$). This means that the $Q$'s
allow to ``linearize'' $f$. This algebra is known to the mathematicians
and was introduced in 1969 by N. Roby (this structure can be generalized
to any polynomial) \cite{cp}.
However, this algebra  is very different from the
usual Clifford algebra. Indeed, ${\cal C}_f$ is defined through $3$-rd order
($n-$th order, in general) 
constraints, and consequently the number of independent monomials
increases with the polynomial's degree 
(for instance, $(Q^1)^2 Q^2$ and $Q^1 Q^2 Q^1$ are 
independent). 
This means that we do not have  enough constraints among the 
generators to order them in some fixed way and, as a consequence,
${\cal C}_f$ turns out to be an infinite dimensional algebra.
However, it has been proved that for any polynomial a finite dimensional
(non-faithful) representation can be obtained \cite{line}. 
But, for polynomial of degree higher  than two, we do
not have  a unique representation,
and  inequivalent representations of ${\cal C}_f$
(even of the same dimension) can be
constructed (see, for instance, \cite{ineq} and below for the 
special cubic polynomials).
Furthermore, the problem of classification of the representations 
of ${\cal C}_f$ is still open, though 
it has been proved that the dimension of the
representation is a multiple of the degree of the polynomial \cite{dimrep}.
For more details one can see \cite{cliffalg} and references therein. 
Thus, the study of the representation of the algebra (\ref{F-poin}) reduces
to a study of the representation of the Clifford algebra of the polynomial
$f(v)=3P.v(v.v)$.\\
  
As mentioned above,
representations of the Clifford algebras of polynomials
are not classified and only some special matrix representation are known. 
For $C_f$ we have found two types of representations.
The first one is constructed with the usual Dirac matrices (and it
can be extended in any space-time dimension),
\beqa
\label{Matbis12}
Q_m = \begin{pmatrix}0&\Lambda^{1/3}\gamma_m&0 \cr
                 0&0& \Lambda^{1/3} \gamma_m \cr
                 \Lambda^{-2/3} P_m&0&0 \end{pmatrix}. & 
\eeqa 

\noindent
The second representation is obtained by linearizing firstly 
the polynomial $P.v((v^0)^2 - (v^3)^2)- P.v((v^1)^2 + (v^2)^2 )$, 

\beqa
\label{Mat9}
&&v.P((v^0)^2 - (v^3)^2) -v.P((v^1)^2+(v^2)^2)
 )= \\
&&\hskip -.9truecm
\begin{pmatrix}0&\Lambda^{-2/3} v.P&0\cr
0&0& \Lambda^{1/3}(v^0+v^3)\cr  \Lambda^{1/3}(v^0-v^3)&0&0\end{pmatrix}^3
 +
\begin{pmatrix}0&\Lambda^{-2/3} v.P&0\cr
0&0& \Lambda^{1/3}(-v^1+i v^2)\cr  \Lambda^{1/3}(v^1+i v^2)&0&0\end{pmatrix}^3
\nonumber
\eeqa

\noindent
with subsequent linearization of this sum of perfect cubes 
by means of the twisted tensorial product \cite{ineq}. 
Similar matrices also appear in a different context in \cite{trirac}
where the cubic root of the Klein-Gordon equation is  studied. The first
representation is $12-$dimensional and the second representation
is  $9-$dimensional.
It is  interesting to notice that in the above two matrix representations,
in order to have matrix elements of the same dimension, a
parameter $\Lambda$  with mass dimension naturally  appears. This means that 
the 3SUSY extension we are studying automatically contains a mass parameter.  

By definition of the algebra (\ref{F-poin2}), the $Q$'s 
are in the vector representation of $\mathfrak{so}(1,3)$. This means 
that $SO(1,3)$ is an outer automorphism of the 3SUSY algebra.
A natural question we should address is whether this automorphism
is  an inner automorphism. When we have
an inner automorphism,  this enables us to write
down the Lorentz transformations (specified by the matrix $\Lambda$) as
$\Lambda^m_{ \ \  n} Q^n = S(\Lambda) Q^m S^{-1}(\Lambda)$.
At the infinitesimal level this reduces to the possibility of  finding
the generators $J_{m n}$ such that
\beqa
\label{eq:inner}
 [J_{m n}  ,Q_r  ]  = \eta_{n r} Q_n -
\eta_{m r} Q_m.
\eeqa

\noindent
For the second representation (\ref{Mat9}), it can
can be  directly checked that there does not exist any $9\times 9$ matrix 
$J_{m n}$ satisfying (\ref{eq:inner}). Therefore, for this representation 
$SO(1,3)$ is an outer automorphism. Thus this representation
breaks down Lorentz invariance. This problem was  also encountered in
\cite{trirac}. In contrast, for the first representation (\ref{Mat9}),
it is easy to see that

\beqa
\label{Lorentz2}
J_{mn} = \frac{1}{4}\left(\gamma_m \gamma_n - \gamma_n \gamma_m\right)+
i(x_m P_n -x_n P_m)
\eeqa

\noindent
with $P_m= -i \frac{\partial}{\partial x_m}$
are the appropriate Lorentz generators acting on the $Q$'s.

\section{Conventions and useful relations}
\renewcommand{\theequation}{B.\arabic{equation}}   
\setcounter{equation}{0} 
In this appendix, we collect useful relations and conventions used in this
paper.
The metric is taken to be 

\beqa
\eta_{m n} = \mathrm{diag}(1,-1,-1,-1)
\eeqa

\noindent
The Levi-Civita tensors $\varepsilon_{mnpq}$  and 
$\varepsilon^{mnpq}=\varepsilon_{rstu} \eta^{mr} \eta^{ns} \eta^{pt} \eta^{qu}$
are  normalized as follow
\beqa
\varepsilon_{0123}=1, \ \ \varepsilon^{0123}= -1
\eeqa

\noindent
In the $\mathfrak{sl}(2,\mathbb C)$ notations of dotted and
undotted indices for two-dimensional spinors, the spinor
conventions to raise/lower indices are as follow (we have minor differences
as compared to the notations of Wess and Bagger \cite{wb}):
$\psi_\alpha =\varepsilon_{\alpha\beta}\psi^\beta$,
$\psi^\alpha =\varepsilon^{\alpha\beta}\psi_\beta$,
$\bar\psi_{\dot\alpha}=\varepsilon_{\dot\alpha
\dot\beta}\bar\psi^{\dot\beta}$, $\bar\psi^{\dot\alpha}
=\varepsilon^{\dot\alpha\dot\beta}\bar\psi_{\dot\beta}$
with $(\psi_\alpha)^* =\bar\psi_{\dot\alpha}$, 
$\varepsilon_{12} = \varepsilon_{\dot 1\dot 2}=-1$,
$\varepsilon^{12} = \varepsilon^{\dot 1\dot 2}=1$.

\noindent
The $4D$
Dirac matrices,  in the Weyl representation,  are 
\begin{eqnarray}
\label{eq:gamma}
\gamma_m =
 \begin{pmatrix}
 0&\sigma_m\\
 \bar\sigma_m&0
 \end{pmatrix},
\end{eqnarray}
with
\begin{eqnarray}\label{eq:spin_rep}
 \sigma_{m\,\alpha\dot\alpha}=\Bigl(1,\sigma_i \Bigr),
 \qquad\qquad\bar\sigma_m{}^{\dot\alpha\alpha}=
 \Bigl(1,-\sigma_i\Bigr),
\end{eqnarray}

\noindent
where the $\sigma_i$'s, $i=1,2,3$, are the Pauli matrices.
The following relation holds,

\beqa
\bar\sigma_m{}^{\dot\alpha\alpha}=
\sigma_{m\,\beta \dot\beta} \varepsilon^{\alpha \beta} 
\varepsilon^{\dot \alpha \dot \beta}.
\eeqa

\noindent
Furthermore, the Lorentz generators for the spinors representation are given 
by

\beqa
\sigma_{mn}{}_\alpha{}^\beta = \frac{1}{4}\left(
\sigma_m{}_{\alpha \dot \alpha} \bar \sigma_n{}^{\dot \alpha \beta}-
\sigma_n{}_{\alpha \dot \alpha} \bar \sigma_m{}^{\dot \alpha \beta}
\right) \nonumber \\
\bar \sigma_{mn}{}^{\dot \alpha}{}_{\dot \beta} = \frac{1}{4}\left(
\bar \sigma_m{}^{\dot \alpha  \alpha}  \sigma_n{}_{ \alpha \dot \beta}-
\bar \sigma_n{}^{\dot \alpha  \alpha}  \sigma_m{}_{ \alpha \dot \beta}
\right)
\eeqa

\noindent
We adopt the usual spinor summation convention

\beqa
\psi \lambda = \psi^\alpha \lambda_\alpha, \ \ 
\bar \psi \bar \lambda = \bar \psi_{\dot \alpha} \bar \lambda^{\dot \alpha}
\eeqa

\noindent
leading to the following Fierz rearrangement:

\beqa
\psi \lambda &=&  \lambda \psi \nonumber \\
\bar \psi \bar \lambda& =&  \bar \lambda \bar \psi \nonumber \\
\bar \psi \bar \sigma_m \lambda& =& -  \lambda \sigma_m \bar \psi \\
\psi \sigma_{mn} \lambda &=& - \lambda \sigma_{mn} \psi \nonumber \\
\bar \psi \bar \sigma_{mn} \bar \lambda& =& 
- \bar \lambda \bar \sigma_{mn} \bar \psi \nonumber
\eeqa

\noindent
With our convention for the Levi-Civita tensor, we have

\beqa
\frac{1}{2} \varepsilon_{mnpq} \sigma^{pq}=i \sigma_{mn} \nonumber \\
\frac{1}{2} \varepsilon_{mnpq} \bar \sigma^{pq}=-i \bar \sigma_{mn}
\eeqa

\noindent
that is,  $\sigma_{pq}$ is a self-dual two-form and
 $\bar \sigma_{pq}$ is an antiself-dual two-form.

It can also be observed that

\beqa
\sigma_{mn}{}^{\alpha \beta} = \sigma_{mn}{}_\gamma{}^\beta 
\varepsilon^{\alpha \gamma} \nonumber \\
\bar \sigma_{mn}{}^{\dot \alpha \dot \beta} = \bar 
\sigma_{mn}{}^{\dot \alpha}{}_{\dot \gamma} 
\varepsilon^{\dot \beta \dot \gamma}
\eeqa

\noindent
are both symmetric in their spinorial indices.

Moreover, We have the following identities

\beqa
\mathrm{Tr}\left(\sigma_{mn} \sigma_{pq}\right) &=&
-\frac{1}{2}\left(\eta_{mp} \eta_{nq}-\eta_{mq} \eta_{np}\right)
+\frac{i}{2} \varepsilon_{mnpq} \\
\mathrm{Tr}\left(\bar \sigma_{mn} \bar \sigma_{pq}\right) &=&
-\frac{1}{2}\left(\eta_{mp} \eta_{nq}-\eta_{mq} \eta_{np}\right)
-\frac{i}{2} \varepsilon_{mnpq} \nonumber 
\eeqa

Starting from Weyl spinors, a Majorana bispinor is given by

\beqa
\psi_M= \begin{pmatrix} \psi_\alpha \cr \bar \psi^{\dot \alpha} \end{pmatrix}
\eeqa

\noindent
and satisfies the Majorana condition

\beqa
\psi_M^c = \psi_M = {\cal C}\bar  \psi_M^t,
\eeqa

where the charge conjugation matrix is defined by  

\beqa
{\cal C} = \begin{pmatrix} {\cal C}_{\alpha \beta} & 0 \cr
                           0& {\cal C}^{\dot \alpha \dot \beta} \end{pmatrix}
         =\begin{pmatrix} \varepsilon_{\alpha \beta} & 0 \cr
                          0& \varepsilon^{\dot \alpha \dot \beta}
 \end{pmatrix}
\eeqa

\noindent
Finally the 3SUSY transformations for  Majorana fermions read

\beqa
\begin{array}{ll}
\delta_\varepsilon \psi_1 = \Lambda^{1/3}\varepsilon^n \gamma_n \psi_2, &
\delta_\varepsilon \bar \psi_1 = -\Lambda^{1/3}\bar \psi_2 \varepsilon^n 
\gamma_n,  \cr
\delta_\varepsilon \psi_2 =\Lambda^{1/3} \varepsilon^n \gamma_n \psi_3, &
\delta_\varepsilon \bar \psi_2 = -\Lambda^{1/3}\bar \psi_3 \varepsilon^n
 \gamma_n,  \cr
\delta_\varepsilon \psi_3 =\Lambda^{-2/3} \varepsilon^n P_n \psi_1, &
\delta_\varepsilon \bar \psi_3 =\Lambda^{-2/3} \varepsilon^n P_n \bar \psi_1. 
\end{array}
\eeqa

\noindent

\end{document}